\documentclass{elsart_modified}
\usepackage{epsfig}
    \newcommand{\ba}{\begin{eqnarray}}
    \newcommand{\ea}{\end{eqnarray}}
    \newcommand{\be}{\begin{equation}}
    \newcommand{\ee}{\end{equation}}

    \newcommand {\bk} {{\mathbf k}}
    
    \newcommand{\AmS}{{\protect\the\textfont2
  A\kern-.1667em\lower.5ex\hbox{M}\kern-.125emS}}

\begin{document}
\runauthor{Junhua and Chuan}
\begin{frontmatter}

\title{Tuning the Tadpole Improved Clover Wilson Action%
      \ on Coarse Anisotropic Lattices\thanksref{fund}}
\author[Beijing]{Junhua Zhang}
\author[Beijing]{and Chuan Liu}
\address[Beijing]{Department of Physics\\
          Peking University\\
                  Beijing, 100871, P.~R.~China}
\thanks[fund]{This work is supported by National Natural
Science Foundation of China (NSFC) and Pandeng fund.}

\begin{abstract}
Wilson quark action, with tadpole improved clover term added, is
studied on coarse anisotropic lattices. The bare velocity of light
parameter in this action is determined non-perturbatively using
the pseudo-scalar and vector meson dispersion relations for
various values of the gauge coupling $\beta$ and bare quark mass
parameter $\kappa$.
\end{abstract}
\end{frontmatter}


\section{Introduction}

 It has become clear that anisotropic, coarse lattices and improved
 lattice actions are the ideal candidates for lattice
 QCD calculations on small computers %
 \cite{lepage93:tadpole,lepage95:pc,colin97,colin99}. They are
 particularly advantageous for heavy objects like the glueballs,
 one meson states with nonzero three momenta and
 multi-meson states with or without three momenta. The gauge action
 employed is the tadpole improved gluonic action on asymmetric
 lattices:
 \ba S=&-&\beta\sum_{i>j} \left[
  {5\over 9}{TrP_{ij} \over \xi u^4_s}
 -{1\over 36}{TrR_{ij} \over \xi u^6_s}
-{1\over 36}{TrR_{ji} \over \xi u^6_s} \right] \nonumber \\
&-&\beta\sum_{i} \left[ {4\over 9}{\xi TrP_{0i} \over  u^2_s}
-{1\over 36}{\xi TrR_{i0} \over u^4_s} \right] \;\;£¬
 \ea
 where $P_{ij}$ is the usual plaquette variable and
 $R_{ij}$ is the $2\times 1$ Wilson loop on the lattice.
 The parameter $u_s$, which we take to be the $4$-th root
 of the average spatial plaquette value, incorporates the
 so-called tadpole improvement and $\xi$ designates the (bare)
 aspect ratio of the anisotropic lattice.

 Using this action, glueball and hadron spectrum has been studied
 within the quenched approximation \cite{colin97,colin99,%
 chuan01:gluea,chuan01:glueb,chuan01:canton1,chuan01:canton2,chuan01:india}.
 The next step would be
 to extend the calculation to more complex physical quantities.
 However, in order to utilize the
 improved fermionic
 action \cite{klassen97:wilson_quark,klassen99:aniso_wilson,aoki01}
 on anisotropic lattices, some parameters
 in the action have to be determined, either perturbatively or
 non-perturbatively, in order to obtain as much improvement as possible.
 In this letter, we will discuss the tuning of
 these parameters in a quenched calculation using clover-improved Wilson
 fermions on anisotropic lattices.
 This paper is organized in the following manner. In Section 2, the
 clover-improved Wilson fermion action on anisotropic lattices is
 introduced. In Section 3,
 we focus on the non-perturbative tuning of the bare
 velocity of light, using the dispersion relations of
 pseudo-scalar and vector mesons. The optimal values of
 the bare velocity of light are obtained for various
 values of $\beta$ and $\kappa$ on different lattices.
 In Section 4, we will conclude with some general remarks.

\section{
Improved Wilson Fermions on Asymmetric Lattices}
\label{sec:fermion}

 Consider a finite four-dimensional lattice with lattice spacing
 $a_\mu$ along the $\mu$ direction for $\mu=0,1,2,3$.
 For definiteness, we also
 denote $a_0=a_t$ and $a_i=a_s$ for $i=1,2,3$. We will
 use $\xi = a_s/a_t$ to denote
 the bare aspect ratio of the asymmetric lattice.
 The un-improved Dirac-Wilson operator on
 asymmetric lattices is given by:
 \be D_0 = {1 \over 2}\sum_{\mu}\nu_\mu
 [(\nabla_\mu +\nabla^*_\mu)\gamma_\mu
 -a_\mu\nabla_\mu\nabla^*_\mu ] \;\;,
 \ee
 where $\nabla_\mu$ and $\nabla^*_\mu$ are the usual
 lattice forward and backward derivatives. The parameter
 $\nu_\mu$ enters the
 Dirac-Wilson operator since the hyper-cubic symmetry is broken. In
 particular, we have $\nu_0=1$ and $\nu_i \equiv \nu$ which we
 call the ``bare'' velocity of light. The clover-improved
 Dirac-Wilson operator is obtained by adding a
 Scheikholeslami-Wohlert, or clover, term to the
 usual Wilson fermion action \cite{sheik-wohlert}.
 To implement the tadpole improvement,
 one replaces each spatial link
 $U_i(x)$ by $U_i(x)/u_s$ while keeping the temporal link
 unchanged. In quenched calculations, one
 usually needs to calculate the quark propagators at various
 valance quark masses. This amounts to different values of $m_0$ or
 $\kappa$ for the same gauge field configuration. In this case, it
 is convenient to use the following fermion matrix:
 \ba
 {\mathcal M}_{xy} &=&\delta_{xy}\sigma + {\mathcal A}_{xy}
 \nonumber \\
 {\mathcal A}_{xy} &=&\delta_{xy}\left[1/(2\kappa_{max})
 +\rho_t \sum^3_{i=1} \sigma_{0i} {\mathcal F}_{0i}
 +\rho_s (\sigma_{12}{\mathcal F}_{12} +\sigma_{23}{\mathcal F}_{23}
 +\sigma_{31}{\mathcal F}_{31})\right]
 \nonumber \\
 &-&\sum_{\mu} \eta_{\mu} \left[
 (1-\gamma_\mu) U_\mu(x) \delta_{x+\mu,y}
 +(1+\gamma_\mu) U^\dagger_\mu(x-\mu) \delta_{x-\mu,y}\right] \;\;,
 \ea
 where the coefficients are given by:
 \ba
 \eta_i &=&\nu/(2u_s) \;\;, \;\;
 \eta_0=\xi/2 \;\;, \;\;\sigma=1/(2\kappa)-1/(1\kappa_{max})\;\;,
 \nonumber \\
 \rho_t &=& c_{SW}(1+\xi)/(4u^2_s) \;\;, \;\;
 \rho_s = c_{SW}/(2u^4_s) \;\;.
 \ea
 In this notation, the bare quark mass
 dependence is singled out into the parameter $\sigma$ and the
 matrix ${\mathcal A}$ remains unchanged if the bare quark mass is
 varied.  Therefore, one could utilize the
 shifted structure of the matrix ${\mathcal M}$ to solve for
 quark propagators at various
 values of $m_0$ (or equivalently $\kappa$)
 at the cost of solving only one value of $\kappa=\kappa_{max}$,
 using the so-called  Multi-Mass Minimal
 Residual ($M^3R$) algorithm %
 \cite{frommer95:multimass,glaessner96:multimass,beat96:multimass}.

 \section{Tuning of the parameters}

 Tuning the bare parameters of the fermion matrix to obtain the
 best improvement is a rather delicate issue
 \cite{klassen99:aniso_wilson}. In a quenched calculation, the
 matter is  simplified since the anisotropy $\xi$ can be
  determined from the pure gauge sector.
  In fact, with the tadpole improvement, the
 renormalization effects of the aspect ratio can be ignored
 \cite{colin97,colin99}. The
 parameter $c_{SW}$ is the hardest to be tuned. In principle, one
 could utilize the Schr\"odinger representation of the QCD path
 integral to determine this parameter non-perturbatively.
 This has been done for the clover-improved Wilson fermion action on
 symmetric lattices \cite{chuan96:non,karl98:csw}. In this work,
 the tadpole improved tree-level value is used.
 In what follows, we will focus on the tuning of
 the parameter $\nu$.
 Since this parameter has rather weak
 dependence on the quark mass on
 anisotropic lattices \cite{klassen99:aniso_wilson}, it can be
 determined non-perturbatively by the pseudo-scalar meson
 dispersion relation for a rather heavy pion.

 Results of this tuning process
 for various values of $\beta$, together with the
 corresponding simulation parameters
 \begin{table}[tb]
 \caption{Simulation parameters for $\beta=1.7$ lattices.
 Input aspect ratio parameter $\xi$ is fixed to be $5$
 throughout this paper.
 The approximate spatial lattice
 spacing $a_s$ in physical units as obtained from
 \cite{colin99} is also shown.
 Results for pion mass, rho mass in lattice
 units, the slope of the pseudo-scalar and vector
 meson dispersion relation
 $Z_{\pi}$ and $Z_{\rho}$ (see Eq.~(\ref{eq:dis}) for
 the definition of them), and the corresponding value
 of $\nu$ and $\kappa$ are also listed.
 \label{table:b17}}
 \begin{tabular}{|c@{\hspace{1mm}}|c@{\hspace{1mm}}c@{\hspace{1mm}}%
 c@{\hspace{1mm}}c@{\hspace{1mm}}||%
 c@{\hspace{1mm}}|c@{\hspace{1mm}}c@{\hspace{1mm}}%
 c@{\hspace{1mm}}c@{\hspace{1mm}}|}
 \hline
 \multicolumn{10}{|c|}{$\beta=1.7$, $a_s\sim 0.39fm$, %
 $6^340$ lattices}
 \\ \hline
 \multicolumn{5}{|c||}{$\nu=0.90$,  $288$ configurations} &
 \multicolumn{5}{c|}{$\nu=0.92$,  $192$ configurations} \\
 \hline
 $\kappa$ & $m_\pi a_t$ & $m_\rho a_t$ & $Z_\pi$ & $Z_\rho$ &
 $\kappa$ & $m_\pi a_t$ & $m_\rho a_t$ & $Z_\pi$  & $Z_\rho$
 \\ \hline
 $.0585$ & $0.3029(8)$ & $0.416(2)$ & $0.97(2)$  & $1.01(5)$ &
 $.059$ &  $0.230(1)$ & $0.360(5)$ & $0.97(2)$  & $1.04(6)$
 \\
 $.0575$ & $0.3803(7)$ & $0.477(1)$ & $1.06(1)$  & $1.08(4)$ &
 $.058$ & $0.315(1)$ & $0.427(2)$ & $1.02(3)$  & $1.12(5)$
 \\
 $.0565$ & $0.4472(5)$ & $0.537(1)$ & $1.09(2)$  & $1.15(4)$ &
 $.057$ & $0.3916(9)$ & $0.491(4)$ & $1.11(2)$  & $1.19(5)$
 \\
 $.0555$ & $0.5159(4)$ & $0.597(1)$ & $1.18(1)$  & $1.06(4)$ &
 $.056$ & $0.4625(8)$ & $0.552(3)$ & $1.17(2)$  & $1.10(7)$
 \\
 $.0545$ & $0.5827(4)$ & $0.657(1)$ & $1.23(1)$  & $1.08(4)$ &
 $.055$ & $0.5319(7)$ & $0.612(2)$ & $1.22(2)$  & $1.15(5)$
 \\ \hline\hline
 \end{tabular}
 \end{table}
 are given in Table~\ref{table:b17} to Table~\ref{table:b30}.
 The value of $\beta$ in our simulations ranges between $1.7$ and $3.0$,
 which roughly corresponds
 to physical spatial lattice spacing $a_s$ in the range
 $0.12\sim 0.4fm$. The aspect ratios of these lattices
 are fixed at $\xi_0=5$. The spatial sizes
 of the lattices range from $L=4$ to $L=8$, which in
 physical units lie in the range $0.5\sim 3.2fm$, depending
 on the value of $\beta$. For simplicity, only the results
 for $6^340$ lattices that are
 close to the optimal value of $\nu$ are shown.
 \begin{table}[tb]
 \caption{Same as Table~\ref{table:b17} except for $\beta=2.2$.
 \label{table:b22}}
 \begin{tabular}{|c@{\hspace{1mm}}|c@{\hspace{1mm}}c@{\hspace{1mm}}%
 c@{\hspace{1mm}}c@{\hspace{1mm}}||%
 c@{\hspace{1mm}}|c@{\hspace{1mm}}c@{\hspace{1mm}}%
 c@{\hspace{1mm}}c@{\hspace{1mm}}|}
 \hline
 \multicolumn{10}{|c|}{$\beta=2.2$, $a_s\sim 0.27fm$, %
 $6^340$ lattices}
 \\ \hline
 \multicolumn{5}{|c||}{$\nu=0.90$, $160$ configurations} &
 \multicolumn{5}{c|}{$\nu=0.95$, $208$ configurations} \\
 \hline
 $\kappa$ & $m_\pi a_t$ & $m_\rho a_t$ & $Z_\pi$ & $Z_\rho$ &
 $\kappa$ & $m_\pi a_t$ & $m_\rho a_t$ & $Z_\pi$  & $Z_\rho$
 \\ \hline
 $.061$ & $0.198(1)$ & $0.291(3)$ & $0.89(4)$  & $0.93(6)$ &
 $.060$ & $0.203(1)$ & $0.294(2)$ & $0.99(4)$  & $0.98(6)$
 \\
 $.060$ & $0.268(1)$ & $0.341(2)$ & $0.96(3)$  & $1.00(5)$ &
 $.059$ & $0.276(1)$ & $0.351(2)$ & $1.04(3)$  & $1.07(5)$
 \\
 $.059$ & $0.3361(7)$ & $0.397(2)$ & $1.00(3)$  & $1.07(4)$ &
 $.058$ & $0.3443(6)$ & $0.406(2)$ & $1.10(2)$  & $1.17(4)$
 \\
 $.058$ & $0.3994(5)$ & $0.452(2)$ & $1.07(2)$  & $1.16(4)$ &
 $.057$ & $0.4091(5)$ & $0.463(1)$ & $1.18(2)$  & $1.25(4)$
 \\
 $.057$ & $0.4616(5)$ & $0.509(1)$ & $1.19(2)$  & $1.18(5)$ &
 $.055$ & $0.4728(4)$ & $0.520(1)$ & $1.26(2)$  & $1.34(3)$
 \\ \hline\hline
 \end{tabular}
 \end{table}
 \begin{table}[tb]
 \caption{Same as Table~\ref{table:b17} except for $\beta=2.4$.
 \label{table:b24}}
 \begin{tabular}{|c@{\hspace{1mm}}|c@{\hspace{1mm}}c@{\hspace{1mm}}%
 c@{\hspace{1mm}}c@{\hspace{1mm}}||%
 c@{\hspace{1mm}}|c@{\hspace{1mm}}c@{\hspace{1mm}}%
 c@{\hspace{1mm}}c@{\hspace{1mm}}|}
 \hline
 \multicolumn{10}{|c|}{$\beta=2.4$, $a_s\sim 0.22fm$, %
 $6^340$ lattices}
 \\ \hline
 \multicolumn{5}{|c||}{$\nu=0.90$, $192$ configurations} &
 \multicolumn{5}{c|}{$\nu=0.92$, $192$ configurations} \\
 \hline
 $\kappa$ & $m_\pi a_t$ & $m_\rho a_t$ & $Z_\pi$ & $Z_\rho$ &
 $\kappa$ & $m_\pi a_t$ & $m_\rho a_t$ & $Z_\pi$  & $Z_\rho$
 \\ \hline
 $.061$ & $0.207(1)$ & $0.265(3)$ & $0.96(3)$  & $1.05(5)$ &
 $.0605$ & $0.204(1)$ & $0.265(3)$ & $0.99(3)$  & $1.10(6)$
 \\
 $.060$ & $0.2713(9)$ & $0.320(2)$ & $1.08(2)$  & $1.13(4)$ &
 $.0595$ & $0.273(1)$ & $0.319(2)$ & $1.08(3)$  & $1.17(4)$
 \\
 $.059$ & $0.3351(8)$ & $0.377(1)$ & $1.16(2)$  & $1.19(3)$ &
 $.0585$ & $0.338(1)$ & $0.376(2)$ & $1.16(3)$  & $1.22(4)$
 \\
 $.058$ & $0.3971(7)$ & $0.434(1)$ & $1.22(2)$  & $1.26(3)$ &
 $.0575$ & $0.3973(8)$ & $0.433(1)$ & $1.26(3)$  & $1.28(4)$
 \\
 $.057$ & $0.4579(6)$ & $0.491(1)$ & $1.28(2)$  & $1.27(3)$ &
 $.0565$ & $0.4587(6)$ & $0.491(1)$ & $1.31(2)$  & $1.31(4)$
 \\ \hline\hline
 \end{tabular}
 \end{table}
 \begin{table}[tb]
 \caption{Same as Table~\ref{table:b17} except for $\beta=2.6$.
 \label{table:b26}}
 \begin{tabular}{|c@{\hspace{1mm}}|c@{\hspace{1mm}}c@{\hspace{1mm}}%
 c@{\hspace{1mm}}c@{\hspace{1mm}}||%
 c@{\hspace{1mm}}|c@{\hspace{1mm}}c@{\hspace{1mm}}%
 c@{\hspace{1mm}}c@{\hspace{1mm}}|}
 \hline
 \multicolumn{10}{|c|}{$\beta=2.6$, $a_s\sim 0.19fm$, %
 $6^340$ lattices}
 \\ \hline
 \multicolumn{5}{|c||}{$\nu=0.85$, $192$ configurations} &
 \multicolumn{5}{c|}{$\nu=0.88$, $192$ configurations} \\
 \hline
 $\kappa$ & $m_\pi a_t$ & $m_\rho a_t$ & $Z_\pi$ & $Z_\rho$ &
 $\kappa$ & $m_\pi a_t$ & $m_\rho a_t$ & $Z_\pi$  & $Z_\rho$
 \\ \hline
 $.062$ & $0.206(1)$ & $0.243(2)$ & $0.99(3)$  & $1.08(4)$ &
 $.0615$ & $0.199(1)$ & $0.236(2)$ & $1.02(3)$  & $1.09(5)$
 \\
 $.061$ & $0.2687(8)$ & $0.301(1)$ & $1.07(2)$  & $1.16(3)$ &
 $.0605$ & $0.2628(8)$ & $0.295(2)$ & $1.11(2)$  & $1.21(3)$
 \\
 $.060$ & $0.3288(6)$ & $0.358(1)$ & $1.16(2)$  & $1.22(3)$ &
 $.0595$ & $0.3243(6)$ & $0.354(1)$ & $1.18(2)$  & $1.22(3)$
 \\
 $.059$ & $0.3886(8)$ & $0.415(1)$ & $1.20(2)$  & $1.26(3)$ &
 $.0585$ & $0.3852(8)$ & $0.411(1)$ & $1.25(2)$  & $1.32(3)$
 \\
 $.058$ & $0.4468(7)$ & $0.471(1)$ & $1.25(2)$  & $1.30(2)$ &
 $.0575$ & $0.4446(7)$ & $0.469(1)$ & $1.31(2)$  & $1.36(3)$
 \\ \hline\hline
 \end{tabular}
 \end{table}
 \begin{table}[tb]
 \caption{Same as Table~\ref{table:b17} except for $\beta=3.0$.
 \label{table:b30}}
 \begin{tabular}{|c@{\hspace{1mm}}|c@{\hspace{1mm}}c@{\hspace{1mm}}%
 c@{\hspace{1mm}}c@{\hspace{1mm}}||%
 c@{\hspace{1mm}}|c@{\hspace{1mm}}c@{\hspace{1mm}}%
 c@{\hspace{1mm}}c@{\hspace{1mm}}|}
 \hline
 \multicolumn{10}{|c|}{$\beta=3.0$, $a_s\sim 0.12fm$, %
 $6^340$ lattices}
 \\ \hline
 \multicolumn{5}{|c||}{$\nu=0.85$, $192$ configurations} &
 \multicolumn{5}{c|}{$\nu=0.88$, $192$ configurations} \\
 \hline
 $\kappa$ & $m_\pi a_t$ & $m_\rho a_t$ & $Z_\pi$ & $Z_\rho$ &
 $\kappa$ & $m_\pi a_t$ & $m_\rho a_t$ & $Z_\pi$  & $Z_\rho$
 \\ \hline
 $.0635$ & $0.149(2)$ & $0.152(2)$ & $1.00(3)$  & $0.90(4)$ &
 $.062$ & $0.191(1)$ & $0.202(2)$ & $1.05(2)$  & $1.09(3)$
 \\
 $.0625$ & $0.199(1)$ & $0.206(2)$ & $1.02(2)$  & $1.09(3)$ &
 $.061$ & $0.246(1)$ & $0.258(1)$ & $1.14(2)$  & $1.16(3)$
 \\
 $.0615$ & $0.253(1)$ & $0.262(1)$ & $1.10(2)$  & $1.16(2)$ &
 $.060$ & $0.302(1)$ & $0.314(1)$ & $1.21(2)$  & $1.24(2)$
 \\
 $.0605$ & $0.3086(9)$ & $0.318(1)$ & $1.17(2)$  & $1.22(2)$ &
 $.059$ & $0.3593(9)$ & $0.371(1)$ & $1.20(2)$  & $1.29(2)$
 \\
 $.0595$ & $0.3644(8)$ & $0.373(1)$ & $1.16(2)$  & $1.27(2)$ &
 $.058$ & $0.4161(8)$ & $0.427(1)$ & $1.26(2)$  & $1.34(2)$
 \\ \hline\hline
 \end{tabular}
 \end{table}

\begin{figure}[htb]
 \begin{center}
 \includegraphics[height=9.0cm,angle=0]{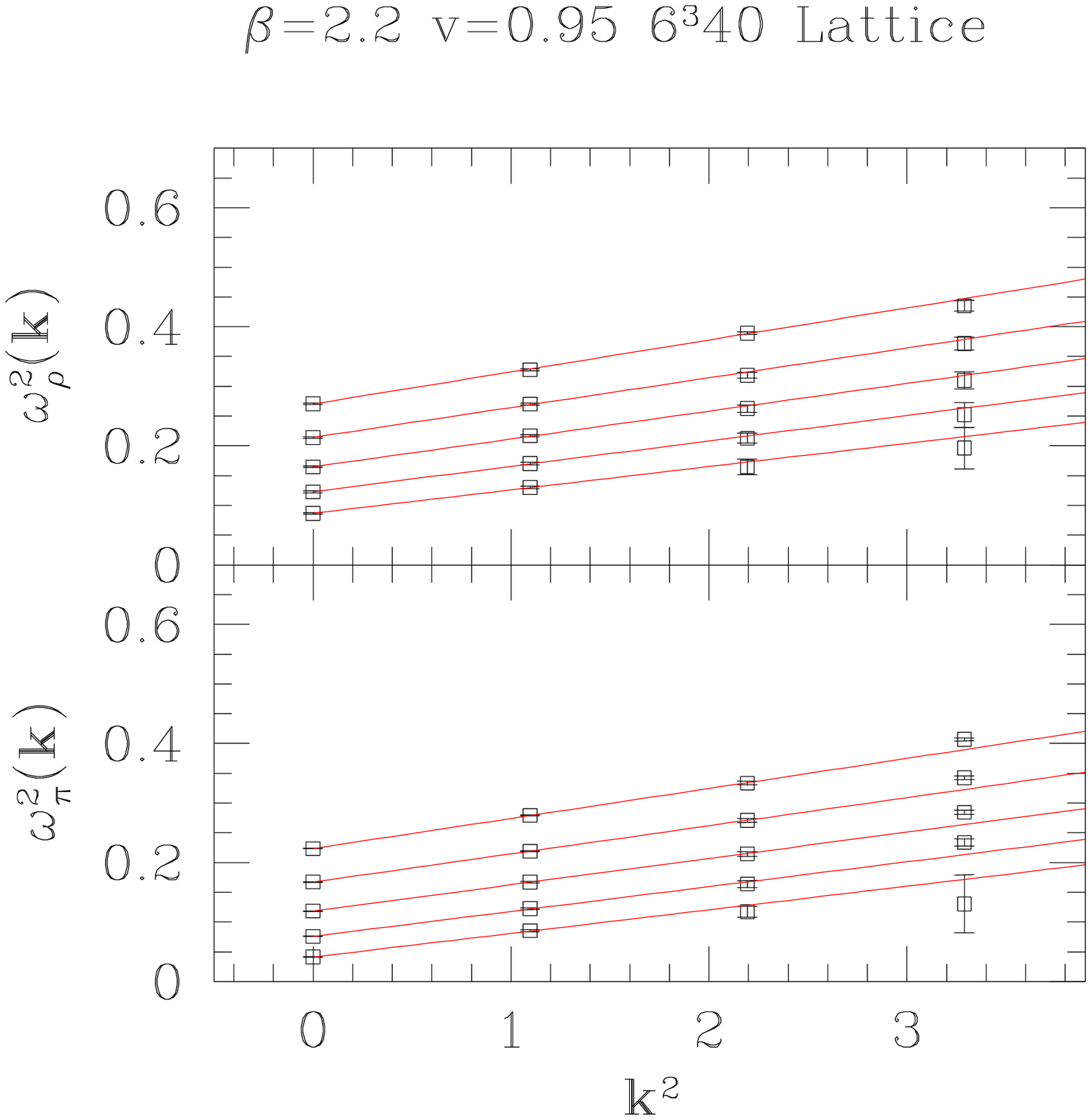}
 \end{center}
 \caption{The dispersion relation for the pseudo-scalar (the lower plot)
  and vector
 meson (the upper plot)
  obtained on $6^340$ lattices with $\beta=2.2$ at five
  different pion masses. The bare
 velocity of light is taken to be $0.95$. It is seen that
 the measured dispersion relations agree with the free dispersion
 relation at small lattice momenta.}
 \label{fig:line06b22}
 \end{figure}

 Quark propagators with various three momenta
 are obtained using the $M^3R$ algorithm
 with wall sources.
 In this work, only the meson energy for zero
 and the following three non-zero lattice
 momenta are measured, namely $(100)$, $(110)$ and $(111)$
 with Dirichlet boundary conditions applied in the temporal
 direction.
 Then the pseudo-scalar correlation functions and
 the vector meson correlation functions are constructed,
 from which the single pion energy and the
 single rho energy are extracted
 in lattice units. Due to the advantage of the
 anisotropic lattices, the meson energy levels can be
 measured with good accuracy. Errors are obtained using the standard
 jack-knife method. The mass of the pion and rho, are  also listed
 in Table~\ref{table:b24}.
 The single meson energy levels
 are fitted according to the expected dispersion relation:
 \be
 \label{eq:dis}
 \omega^2_\pi(\bk) = m^2_\pi + Z_\pi \bk^2 \;\;,
 \;\;
 \omega^2_\rho(\bk) = m^2_\rho + Z_\rho \bk^2 \;\;.
 \ee
 In Fig.~\ref{fig:line06b22},
 the dispersion relations of the pseudo-scalar
 (the lower plot) and the vector meson (the upper plot) are
  shown for $\beta=2.2$, $\nu=0.95$ and five different values
  of $\kappa$ on $6^340$ lattices. The straight lines
  represent the fits using the first three lowest momentum
  points (including the zero-momentum point)
 according to Eq.~(\ref{eq:dis}).
 It is seen that the above dispersion relations fit
 the simulation data well
 for lattice momenta that are not too large.
 Fitting qualities are similar
 for other parameter sets. The results of  the slope
 obtained from the fitting are also included in Table~\ref{table:b24}.
 By tuning the bare velocity
 of light parameter $\nu$, we obtain a value of
 $Z_\pi$ which is consistent with unity.
 The same tuning procedure is also applied to smaller ($4^340$)
 and larger ($8^340$) lattices. The outcome of the procedure
 is  quite similar. No significant dependence
 of the parameter $\nu$ on the volume of the lattice has
 been observed.

\section{Conclusions}

 In this letter, we outline the method on the non-perturbative
 determination of the bare velocity of light parameter
 $\nu$ in the anisotropic Wilson quark action %
 \footnote{We have just noticed that, in a recent
 work, similar tuning process has been studied by
 H.~Matsufuru et al. \cite{matsufuru}.}.
 The optimal values of $\nu$ are obtained for various
 values of $\beta$ and $\kappa$. These values can then be
 utilized in a  quenched calculation using the improved
 anisotropic Wilson action. Some of these calculations are
 in progress and we hope to report the results soon.


\end{document}